\documentclass[twocolumn,pra,showpacs,floatfix,superscriptaddress]{revtex4}
\usepackage{graphicx}
\usepackage{amsmath}
\usepackage{latexsym} 
\usepackage{dcolumn}
\usepackage{verbatim}
\usepackage[latin1]{inputenc}
\usepackage{wasysym}

\newcommand{\be}{\begin{equation}}
\newcommand{\ee}{\end{equation}}
 
\newcommand{\rr}{{\bf r}}
\newcommand{\A}{{\bf A}}
\newcommand{\C}{{\bf C}}
\newcommand{\M}{{\bf M}}
\newcommand{\T}{{\bf T}}
\newcommand{\W}{W}

\newcommand{\E}{{\mathcal E}}

\newcommand{\ket}[1]{|#1\rangle}

\begin{document}

\advance\voffset by 0.9cm

\title{Relativistic calculations of pionic and kaonic atoms hyperfine structure}

\author{Martino Trassinelli}

\email{m.trassinelli@gsi.de}
\affiliation{Gesellschaft f{ü}r Schwerionenforschung, Darmstadt, Germany}
\affiliation{Laboratoire Kastler Brossel,
{É}cole Normale Sup{é}rieure; CNRS; Universit{é} Pierre et Marie Curie-Paris 6, Paris, France}

\author{Paul Indelicato}

\email{paul.indelicato@lkb.ens.fr}
\affiliation{Laboratoire Kastler Brossel,
{É}cole Normale Sup{é}rieure; CNRS; Universit{é} Pierre et Marie Curie-Paris 6, Paris, France}

\date{\today}

\begin{abstract}
We present the relativistic calculation of the hyperfine structure in pionic and kaonic atoms.
A perturbation method has been applied to the Klein-Gordon equation to take into account the relativistic corrections.
The perturbation operator has been obtained \textit{via} a multipole expansion of the nuclear electromagnetic potential.
The hyperfine structure of pionic and kaonic atoms provide an additional term in the quantum electrodynamics calculation of the energy transition of these systems. Such a correction is required for a recent measurement of the pion mass.
\end{abstract}


\maketitle 

% main text

\section{Introduction}
In the last few years transition energies in pionic  \cite{Gotta2004} and kaonic atoms \cite{Beer2002} have
 been measured with an unprecedented precision. 
The spectroscopy of pionic and kaonic hydrogen allows to study the strong interaction at low energies
 \cite{Gasser1983,Lyubovitskij2000,Gasser2002} by measuring the energy and natural width of the ground
 level with a precision of few meV \cite{ProposalPH,Anagnostopoulos2003a,Beer2005}.
Besides, light pionic atoms can additionally be used to define new low-energy X-ray standards \cite{Anagnostopoulos2003b}
 and to evaluate the pion mass using high accuracy X-ray spectroscopy
 \cite{Lenz1998,ProposalPM,Nelms2002a,MyThesis}. Similar endeavour are in progress with kaonic atoms \cite{Beer2002}.

In this paper we present the calculation of the hyperfine structure in pionic and kaonic atoms considering the perturbation term due to the interaction between the pion or kaon orbital moment with the magnetic momentum of the nucleus. 
Non-relativistic calculations for the pionic atom hyperfine structure can be found in 
Ref.~\cite{Ebersold1978,Koch1980,Scheck}.
Other theoretical predictions for HFS including relativistic corrections can be found only 
for spin-$\frac 1 2$ nucleus \cite{Austen1983,Owen1994}. 
Contrary to these methods, our technique is not restricted to this case and it can be used for an arbitrary value of the nucleus spin including automatically the relativistic effects.
In particular, we calculate the HFS energy splitting for pionic nitrogen, which has been used for a recent measurement of the pion mass aiming at an accuracy of few ppm \cite{Lenz1998,ProposalPM,MyThesis,Trassinelli2007}, and for kaonic nitrogen that has been proposed for the kaon mass measurement \cite{Beer2002}.

This article is organized as follows. In Sec.~\ref{sec:pert} we calculate the first energy correction applying a perturbation method for the Klein-Gordon equation. In Sec.~\ref{sec:HFS} we will  obtain the perturbation term using the multipole expansion of the nuclear electromagnetic potential. Section~\ref{sec:numerical} is dedicated to the numerical calculations for some pionic and kaonic atoms. 

\section{Calculation of the energy correction \label{sec:pert}}
The relativistic dynamic of a spinless particle is described by the Klein-Gordon equation. The electromagnetic interaction between a negatively charged spin-0 particle with a charge equal to $q=-e$ and the nucleus can be taken into account introducing the nuclear potential $A_\nu$ in the KG equation via the minimal coupling $p_\nu \to p_\nu -q A_\nu$ \cite{Bjorken-Drell-RQM}. 
In particular, in the case of a central Coulomb potential $(V_0(r), \boldsymbol{0})$, the KG equation for a particle with a mass $m$ is:
\be
m^2 c^2 \Psi_0(x) = \left\{ \frac {1}{c^2} \left[ i \hbar \partial_t + e V_0(r)\right]^2 + \hbar^2 \nabla^2 \right\} \Psi_0(x),  \label{eq:general}
\ee
where $\hbar$ is the Planck constant, $c$ the velocity of the light and the scalar wavefunction $\Psi_0(x)$ depends on the space-time coordinate $x=(c t,\rr)$. 
We  consider here the stationary solution of Eq.~\eqref{eq:general}. In this case, we can write:
\be
\Psi_0(x)=\exp(-iE_0 t/\hbar)\ \varphi_0(\rr)
\ee
and  Eq.~\eqref{eq:general} becomes:
\be
\left\{ \frac{1}{c^2} \left[ E_0 + e V_0(r) \right]^2 + \hbar^2\nabla^2 - m^2c^2 \right\} \varphi_0(\rr) = 0, \label{eq:zero}
\ee
where $E_0$ is the total energy of the system (sum of the mass energy $m c^2$ and binding energy $\E_0$).

The perturbation correction $E_1$ can be deduced introducing an additional operator $\W$ in the zeroth order equation:
\be
\left\{ \frac{1}{c^2} \left[ E + e V_0(r) \right]^2 + \hbar^2\nabla^2 - m^2c^2 - \W(\rr) \right\} \varphi(\rr) = 0. \label{eq:pert}
\ee
$\W(\rr)$ is in general non-linear. In the case of a correction $V_1$ to the Coulomb potential $V_0$, we have:
\be
\W(\rr) = - \cfrac 1 {c^2} \left( 2 e^2 V_0(\rr) V_1(\rr) + 2 e E\ V_1(\rr) \right)
\ee
If we consider the interaction with the nuclear magnetic field as a perturbation, we have:
\be
W(\rr) = i\hbar e \left[ 2 A_i(\rr) \partial^i + [\partial_i,A^i(\rr)] \right]  - e^2A^i(\rr)A_i(\rr).
\ee

The correction to the energy due to $W$ can be calculated perturbatively with some manipulation of Eqs.~\eqref{eq:zero} and \eqref{eq:pert} \cite{Decker1987,Lee2006}, or \textit{via} a linearization of the KG equation using the Feeshbach-Villars formalism \cite{Friar1980,Leon1981}.
In both cases we have
\be
E_1 = \frac{c^2 \langle \W \rangle}{2 \left( E_{(0)}^{nl} + \langle e V_0 \rangle \right)}. \label{eq:E1-ericson}
\ee
where we define
\be
\left\langle A \right\rangle   = \cfrac {\int_V \varphi^*(\rr) A(\rr,t) \varphi(\rr)\ d^3r}{\int_V \varphi^*(\rr) \varphi(\rr)\ d^3r}.
\ee
Equation \eqref{eq:E1-ericson} is valid for any wavefunction normalization.

\section{Calculation of the hyperfine structure operator\label{sec:HFS}}
%The hyperfine structure in pionic and kaonic atoms arises from the interaction between the nuclear magnetic moment and the orbital magnetic moment of the particle that can be taken into account by introducing a new term in the KG equation.

%This term is obtained from the  multipole development of the nuclear electromagnetic field  \cite{Schwartz1955,Lindgren1974,Cheng1985}.

The expression for $W(\rr)$ in the HFS case is derived using the multipole development of the 
vector potential $\A( \rr )$ in the Coulomb gauge \cite{Schwartz1955,Lindgren1974,Cheng1985}.
We neglect here the effect due to the  the spatial distribution of the nuclear magnetic moment  in the nucleus \cite{Bohr1950} (Bohr-Weisskopf effect), while
effect due to the charge distribution (Bohr-Rosenthal effect) are included in the numerical results of Sec.~\ref{sec:numerical}. 

The hyperfine structure due to the magnetic dipole interaction is obtained by taking into account 
the first magnetic multipole term. Using the Coulomb gauge we have \cite{Schwartz1955,Lindgren1974}:
\be
\A(\rr) = -i \frac {\mu_0}{4 \pi } \sqrt{2} r^{-2}  \C^{11} \circ \M^1. \label{eq:mult_dev}
\ee
where the symbol ``$\circ$'' indicates here the general scalar product between 
tensor operators. $\M^1$  operates only on the nuclear 
part $\ket{ I m_I}$ and $\C^{11}$ is the  vector sperical harmonic \cite{Lindgren1974,Edmonds} acting on the pion part $\ket{n l m }$ of the wavefunction. 
We can decompose the perturbation term $W(\rr)$ as:
\be
W(\rr) = W_1(\rr) + W_2(\rr),
\ee
where
\be
W_1(\rr) = +i\hbar e \left[ 2 A_i(\rr) \partial^i + [\partial_i,A^i(\rr)] \right],
\ee
is the linear part and 
\be
W_2(\rr) = - e^2A^i(\rr)A_i(\rr),
\ee
is the quadratic part.

We study first the operator $W_1$. 
We note that $[\partial_i,A^i(\rr)] = - \boldsymbol{\nabla} \cdot \A(\rr) = 0$ since we are using the Coulomb gauge. In this case
we have:
\begin{multline}
\W_1(\rr) = +2 i \hbar e A_i(\rr) \partial^i = - 2 i \hbar e \A(\rr)\cdot \boldsymbol{\nabla} =  \\
- e \mu_0 \hbar \frac{\sqrt 2}{ 2 \pi } r^{-2} ( \C^{11} \cdot \boldsymbol{\nabla} ) \circ \M^1.
\end{multline}
Using the properties of the spherical tensor \cite{Lindgren1974,Edmonds}, we can show that:
\be
\C^{11}_q \cdot \boldsymbol{\nabla} = -\frac{r^{-1}}{\sqrt 2}  L_q,
\ee
where $ L_q$ is the dimensionless angular momentum operator in spherical coordinates.
The perturbation operator can be written as a scalar product in spherical coordinate of the
operator $\T^1$ acting on the pion wavefunction, and the nuclear operator $\M^1$:
\be
\W_1(\rr) = \frac{ e \mu_0 \hbar }{2 \pi } r^{-3}( \boldsymbol{ L}\circ\M^1) = \T^1 \circ \M^1 
\ee
with
\be  
T_q^1 = \frac{e \mu_0 \hbar}{2 \pi } r^{-3}   L_q. \label{eq:Tq}
\ee

The expected value of the operator $\W_1$ can be evaluated applying the scalar product properties in spherical coordinates \cite{Edmonds,Lindgren-Morrison}:
\begin{multline}
\langle n'l'IF'm_F'| \W_1 | nlIFm_F\rangle = 
(-1)^{l + I + F}\delta_{F F'}\delta_{m_F m_F'}\delta_{I I'} × \\
\left\{\begin{array}{ccc}
F & I & l'\\
1 & l & I
\end{array}\right\} 
\langle n'l' \|  T^1 \| n l\rangle \langle I \|  M^1 \| I\rangle, 
\end{multline}
where $\big\{ \begin{array} {c c c} a & b & c \\ d & e & f \end{array} \big\}$ represents a Wigner 6-j symbol. 
The reduced operator $\langle n'l' \| T^1 \| n l\rangle$ is
calculated from the matrix elements $\langle n'l'm'|   T_q^1 | n l m\rangle$ by a particular choice of the quantum numbers $m$ and $q$ applying the Wigner-Eckart theorem:
\begin{multline}
\langle n'l' \|  T^1 \| n l\rangle = \delta_{l 0}\frac {(-1)^{l-1}}
{\left(\begin{array}{ccc}
l & 1 & l\\
-1 & 0 & 1
\end{array}\right) } \langle n'l'1|   T_0^1 | n l1\rangle =  \\
= \delta_{l 0} \sqrt{l} \sqrt{l + 1} \sqrt{2l +1} \frac{e \mu_0 \hbar}{2 \pi }\langle n'l'1| r^{-3}  L_z | n l1\rangle = \\
  = \delta_{l 0} \delta_{l l'}  \sqrt{l} \sqrt{l + 1} \sqrt{2l +1} \frac{e \mu_0 \hbar}{2 \pi }\langle n'l|  r^{-3} | nl\rangle,  \label{eq:Tq-final}
\end{multline}
where $\big( \begin{array} {c c c} a & b & c \\ d & e & f \end{array} \big)$ indicates the  Wigner 3-j symbol.

The nuclear operator can be related to the magnetic moment on the nucleus
by $\langle I I| M^1_0 | I I\rangle =  \mu_I \mu_N$ \cite{Lindgren1974,Cheng1985} where $\mu_I$ is the nuclear dipole momentum in units of the nuclear magneton $\mu_N = e \hbar /2 m_p c$:
\be
\langle I \|  M^1 \| I\rangle = \frac{ \mu_I \mu_N}
{\left(\begin{array}{ccc}
I & 1 & I\\
-I & 0 & I
\end{array}\right)}. \label{eq:W-E}
\ee

Considering Eq.~\eqref{eq:Tq}, the total expression for $W_1(\rr)$ becomes:
\begin {multline}
\langle n'l'I F'm_F'|   W_1 | nlIFm_F\rangle = \\
= \delta_{F F'}\delta_{m_F m_F'} \delta_{l l'}\mu_I \mu_N × \\
\frac{e \mu_0 \hbar }{2 \pi } \frac{F(F+1) - I(I+1) -l(l+1))}{2 I}\langle n'l|  r^{-3} | nl\rangle. \label{eq:w_rel}
\end {multline}
which, as expected, is equal to zero for $l=0$ (then $I=F$).

To find the final expression of the HFS energy shift, we have to evaluate the contribution of 
the operator  
$W_2 (\rr) =-e^2A^i(\rr)A_i(\rr)$ in the $\langle   W \rangle$ diagonal terms. 
Using Eq.~\eqref{eq:mult_dev}, we have: 
\begin{multline}
\langle n l I F m_F | W_2| n l I F m_F\rangle = 
+ 2 \left( \frac{ e \mu_0}{4 \pi} \right)^2 × \\
\langle n l I F m_F |( r^{-2} \C^{11} \circ \M^1) \cdot ( r^{-2} \C^{11} \circ \M^1)| n l I F m_F\rangle.
\end{multline}
We are in presence of three independent scalar products: two scalar products between the tensor 
$\C^{11}$ and the vector $\M^1$, and the scalar product between the vectorial operators  $\C^{11} \circ \M^1$. 
The ``$\cdot$'' scalar product in $W_2$ can be decomposed using the properties of the 
reduced matrix elements of
a generic operator product $X^K$, of rank $K$, between non-commutating tensor operators $U^k$ and $V^k$
of rank $k$. For our case, this scalar product corresponds 
to a tensor product with $K=0$,  and $k=1$: 
\begin{align}
&X^0= U^1 \cdot  V^1 =
( r^{-2} \C^{11} \circ \M^1) \cdot ( r^{-2} \C^{11} \circ \M^1), \\
&U^1 = V^1 = ( r^{-2} \C^{11} \circ \M^1).
\end{align}
We have \cite{Edmonds}: 
\begin{multline}
\langle n l I F ||  X^0 || n l I F \rangle =  
\sum_{F'}
\left\{\begin{array}{ccc}
1 & 1 & 0\\
F & F & F'
\end{array}\right \} × \\
\langle n l I F ||  U^1 || n l I F' \rangle\langle n l I F' || V^1 || n l I F \rangle.
\end{multline}
$\boldsymbol{V}^1$ and $\boldsymbol{U}^1$ are scalar products between commutative tensor operators, and their 
reduced matrix element $\langle F' ||  V || F \rangle$ can be calculated applying again the Wigner-Eckart theorem for the component $q=0$:
\be
\langle n l I F' ||  V^1 || n l I F \rangle = 
\frac{\langle n l I F' F' |  V_0^1 | n l I F F \rangle}{\left(\begin{array}{ccc}
F' & 1 & F\\
-F' & 0 & F
\end{array}\right)}.
\ee
where \cite{Varshalovich}:
\be
V_0^1 = U_0^1 =  r^{-2} (\C^{11} \circ \M^1)_0 =  r^{-2} \sum_q(-1)^q \frac{q}{\sqrt{2}} C^1_q M^1_{-q}.
\ee
To evaluate the matrix element $\langle n l I F' F' |  V_0^1 | n l I F F \rangle$ we can explicitly decompose $| n l I F F \rangle$ as a function of the eigenfunctions $| n l m\rangle$ and $| I m_I\rangle$ using the Clebsch-Gordan coefficients.
We have
\begin{multline}
\langle n l I F' F' |  V_0^1 | n l I F F \rangle = \\
\frac{-1}{r^2 \sqrt{2}} \sum_{m', m_I', m, m_I} \langle l m' I m_I' | l I F' F' \rangle \langle l m I m_I | l I F F \rangle × \\ 
\big[ \langle n l m' | r^{-2} C^1_1 |n l m \rangle \langle I m_I'| M^1_{-1} | I m_I \rangle -\\
\langle n l m' | r^{-2} C^1_{-1} |n l m \rangle \langle I m_I'| M^1_{1} | I m_I \rangle \big].
\end{multline}
Applying the Wigner-Eckart theorem we obtain
\begin{multline}
\langle n l I F' F' |  V_0^1 | n l I F F \rangle = \\
\frac{-1}{r^2 \sqrt{2}} \langle n l \| r^{-2} C^1 \| n l \rangle \langle I \| M^1 \| I \rangle × \\
\sum_{m', m_I', m, m_I} \langle l m' I m_I' | l I F' F' \rangle \langle l m I m_I | l I F F \rangle × \\
\Bigg[
{\left(\begin{array}{ccc}
l & 1 & l\\
-m' & 1 & m
\end{array}\right)}
{\left(\begin{array}{ccc}
I & 1 & I\\
-m_I' & -1 & m_I
\end{array}\right)}
- \\
{\left(\begin{array}{ccc}
l & 1 & l\\
-m' & -1 & m
\end{array}\right)}
{\left(\begin{array}{ccc}
I & 1 & I\\
-m_I' & +1 & m_I
\end{array}\right)}
\Bigg].
\end{multline}
The reduced matrix element $\langle n l \| r^{-2} C^1 \| n l \rangle$ can be decomposed in a radial and angular part
\be
\langle n l \| r^{-2} C^1 \| n l \rangle = \langle n l | r^{-2} | n l \rangle\langle l \| C^1 \| l \rangle.
\ee
Due to the symmetry properties, $\langle l \| C^1 \| l \rangle$ is equal to zero for any $l$ \cite{Edmonds}.
This result implies that the reduced matrix elements of $U^1$ and $V^1$ are always equal to zero.  
As a consequence, the diagonal elements $\langle A^i(\rr)A_i(\rr)\rangle=0$
for any wavefunction, i.e., $W_2$ does not contribute to the HFS energy shift.  

\hfill

We can now write the final expression for the HFS energy correction:
\begin{multline}
 E_1^{nlF} = \frac{\mu_I \mu_N e \mu_0 \hbar c^2}{4 \pi \left(E_0^{n l} - \langle  n l | V_0(r) |  n l \rangle \right)}  × \\
\left[ \cfrac{F(F+1) - I(I+1) -l(l+1)}{2 I} \right]\langle  n l |  r^{-3} |  n l \rangle. \label{eq:HFS_final_SI}
\end{multline}

This formula is obtained by a perturbation approach of the KG equation.
For this reason, all the relativistic effects are automatically included in Eq.~\eqref{eq:HFS_final_SI}.
In the non-relativistic limit $c \to \infty$, $(E_0 - \langle V \rangle)/c^2 \to m$ and  we find the usual expression of the HFS for the Sch{ö}dinger equation \cite{Bethe-Salpeter}

\section{Numerical results and behaviors\label{sec:numerical}}
We present here some calculations for a selection of pionic and kaonic atom transitions. 
Such calculations are obtained solving numerically the Klein-Gordon equation using the multi-configuration Dirac-Fock code developed by one of the author (P.I.) and J.-P.~Desclaux \cite{Desclaux1975,Desclaux1993,Boucard2000,Desclaux2003} that has been modified to include spin-0 particles case, even in the presence of electrons \cite{Santos2005}.
The first part is dedicated to the $5 \to 4$ and $8 \to 7$ transitions in pionic and kaonic nitrogen, respectively.
In the second part we will study the dependence of the HFS splitting against the nuclear charge $Z$ to observe the role of the relativistic corrections.

\subsection{Calculation of the energy levels of pionic and kaonic nitrogen}  
The precise measurement of $5g \to 4f$ transition in pionic nitrogen and the related QED predictions allow for the precise measurement of the pion mass \cite{Lenz1998,ProposalPM,Nelms2002a,MyThesis,Trassinelli2007}.
In the same way, the transition $8k \to 7i$ in kaonic nitrogen can be used for a precise mass measurement of the kaon \cite{Beer2002}.
For these transitions, strong interaction effects between meson and nucleus are negligible, and the level energies are directly dependent to the reduced mass of the atom.
The nuclear spin of the nitrogen isotope $\rm ^{14}N$ is equal to one leading to the presence of several HFS sublevels.
The observed transition is a combination of several different transitions between these sublevels,
causing a shift  that has to be taken into account to extract the pion mass from the experimental values.
Transition probabilities between HFS sublevels can easily be calculated using the non-relativistic formula \cite{Bethe-Salpeter,LandauEMQ} (the role of the relativistic effects is here negligible), 
if one neglect the HFS contribution to the transition energy:
\begin{multline}
A_{n l I F \to n' \ l' I F'} = \\
\frac{(2 F + 1)(2 F' + 1)}{2 I + 1 } \label{eq:trans-prob}
\left\{\begin{array}{ccc}
l' & F' & I\\
F & l & 1
\end{array}\right \}^2
A_{n l \to n' l'},
\end{multline}
where  
\be
A_{n l \to n' l'} = \frac{4(E_{n l} -  E_{n' \ l - 1})^3}{3 m^2 c^4 \hbar }\frac{\alpha }{(Z \alpha )^2}\frac {l}{2 l + 1}( R^{n' \ l - 1}_{n l})^2,
\ee
with 
\be
R^{n' l'}_{n l} = \frac{1}{a_0^2}\int_0^\infty \phi^*_{n l}(r) \phi_{n' l'}(r) r^3 dr.
\ee
$a_0= \hbar / (m c Z \alpha )$ is the Bohr radius and $\phi_{n l}$ are the non-relativistic wavefunctions.

\begin{table}
\caption{Energy (in eV) contribution for the selected levels in pionic nitrogen. The first error takes into account neglected next order QED corrections. The second is due to the accuracy of the pion mass ($± 2.5$~ppm).
\label{tab:piN}     
}
\begin{tabular}{l r r }     
\hline
\hline                    
&          5g-4f   &       5f-4d   \\
\hline
Coulomb &       4054.1180       &       4054.7189       \\
Finite size     &       0.0000  &       0.0000  \\
Self Energy     &       -0.0001 &       -0.0003 \\
Vac. Pol. (Uehling)     &       1.2485  &       2.9470  \\
Vac. Pol. Wichman-Kroll &       -0.0007 &       -0.0010 \\
Vac. Pol. Loop after Loop       &       0.0008  &       0.0038  \\
Vac. Pol. K{ä}ll{é}n-Sabry  &       0.0116  &       0.0225  \\
Relativistic Recoil     &       0.0028  &       0.0028  \\
HFS Shift       &       -0.0008 &       -0.0023 \\
Total   &       4055.3801       &       4057.6914       \\
\hline
Error & $± 0.0011$ & $± 0.0011$ \\
Error due to the pion mass & $± 0.010$ & $± 0.010$ \\   
\hline
\hline
\end{tabular}
\end{table} 
\begin{table} 
\caption{Hyperfine transition energies and transition rate in pionic nitrogen.
\label{tab:piN_energies} 
} 
\begin{tabular}{c c r r r r } 
\hline 
\hline
Transition      &       F-F'    &       Trans. rate (s$^{-1}$)  &       Trans. E (eV)   &       Shift (eV)           \\     
\hline 
$5f \to 4d$	&	4-3	& $	4.57	\times 10^{13}$	&	4057.6876	&	-0.00606	\\
	&	3-2	& $	3.16	\times 10^{13}$	&	4057.6970	&	0.00341	\\
	&	3-3	& $	2.98	\times 10^{13}$	&	4057.6845	&	-0.00910	\\
	&	2-1	& $	2.13	\times 10^{13}$	&	4057.7031	&	0.00946	\\
	&	2-2	& $	2.25	\times 10^{13}$	&	4057.6948	&	0.00112	\\
	&	2-3	& $	0.01	\times 10^{13}$	&	4057.6822	&	-0.01138	\\
$5g \to 4d$ 	&	5-4	& $	7.13	\times 10^{13}$	&	4055.3779	&	-0.00304	\\
	&	4-3	& $	5.47	\times 10^{13}$	&	4055.3821	&	0.00113	\\
	&	4-4	& $	5.27	\times 10^{13}$	&	4055.3762	&	-0.00482	\\
	&	3-2	& $	4.17	\times 10^{13}$	&	4055.3852	&	0.00420	\\
	&	3-3	& $	0.36	\times 10^{13}$	&	4055.3807	&	-0.00029	\\
	&	3-4	& $	0.01	\times 10^{13}$	&	4055.3747	&	-0.00624	\\

\hline 
\hline
\end{tabular}
\end{table}

For these calculations, presented in Tables~\ref{tab:piN} and \ref{tab:piN_energies}, we used the nitrogen nuclear mass value from Ref.~\cite{Audi2003}. 
The {\itshape Coulomb} term in the Table includes the non-relativistic recoil correction using the reduced mass on the KG equation.
The pion and nucleus charge distribution contribution are also included. The pion charge distribution radius contribution is included following
\cite{Santos2005,BoucardThesis}. For the pion charge distribution radius we take  $r_\pi=0.672 ± 0.08$ \cite{PDBook}. For the nuclei we take values from Ref.~\cite{Angeli2004}.
The leading QED corrections, vacuum polarization, contribution is calculated self-consistently, thus taking into account 
the loop-after-loop contribution to all orders, at the Uehling approximation.
 This is obtained by including the Uehling potential into the KG equation \cite{Boucard2000}. Other Higher-order vacuum
 polarization contribution are calculated as perturbation to the KG equations: {\itshape  Wichman-Kroll } and   {\itshape  K{ä}ll{é}n-Sabry } \cite{Fullerton1976,Huang1976}.  
The self-energy is calculated using the expression in  Ref.~\cite{Jeckelmann1985} and it includes the recoil correction. 
The \textit{Relativistic recoil} term has been evaluated adapting the formulas from Refs.~\cite{Barker1955,Owen1994} (more details can be found in Ref.~\cite{MyThesis}).
\begin{figure}
  \begin{center}
    \includegraphics[width=0.48\textwidth]{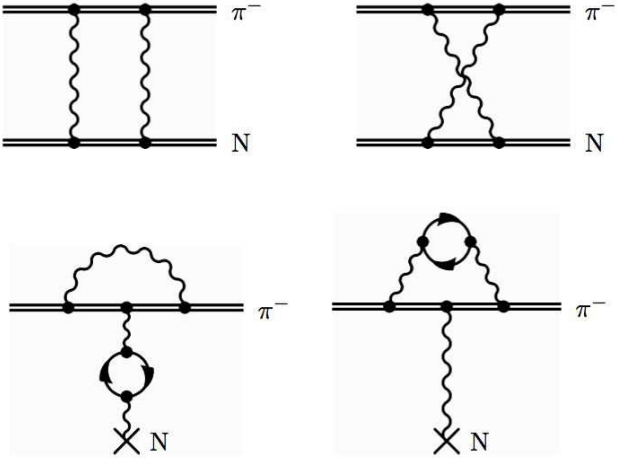}
  \end{center}
  \caption{Diagrams relative to the unevaluated QED contributions: second order recoil correction (top), and vacuum polarization and self-energy mixed diagrams (bottom). Their effects are estimated using the available formulas for spin-$\frac 1 2$ particles.}
  \label{error-terms}
\end{figure}
The calculations presented here do not  take into account second order recoil effects (Fig.~\ref{error-terms} top),
 or higher QED corrections as vacuum polarization and self-energy mixed diagrams (Fig.~\ref{error-terms} botton).
The contribution from these terms has been estimated using the formula for
a spin-$\frac 1 2$ particle with a mass equal to the pion's.
For the $5 \to 4$ pionic nitrogen transitions, vacuum polarization and self-energy mixed diagrams contribute  in 
the order of 1~meV for the diagram with the vacuum polarization loop in the nuclear photon line \cite{Pachucki1996}
 (Fig.~\ref{error-terms} bottom-left),
and 0.0006~meV for the diagram with the vacuum polarization loop inside the self-energy loop \cite{Eides2001}
 (Fig.~\ref{error-terms} bottom-right).
The second order recoil contributions are in the order of 0.04 meV \cite{Sapirstein1990} (Fig.~\ref{error-terms} top).
The largest contribution comes from the unevaluated diagram  
with the vacuum polarization loop in the nuclear photon line \cite{Pachucki1996}.

Assuming a statistical population distribution of the HFS sublevels, we can use Eq.~\eqref{eq:trans-prob} to calculate
 the mean value of the transitions using the results in Table~\ref{tab:piN_energies}. 
Comparing this calculation with the one without the HFS, we obtain a value for the HFS shift. 
For transitions $5g \to 4f$ and $5f \to 4d$ we obtain shifts of 0.8 and 2.2~meV, respectively. 
These values correspond to a correction to the pion mass between 0.2 and 0.6~ppm.

\begin{table}
\caption{Energy (in eV) contribution for the selected levels in kaonic nitrogen. The first error takes into account neglected next order QED corrections. The second is due to the accuracy of the kaon mass ($± 32$~ppm).
\label{tab:kN}     
}
\begin{tabular}{l r r }     
\hline
\hline                    
&          8k-7i   &       8i-7h   \\
Coulomb	&	2968.4565	&	2968.5237	\\
Finite size	&	0.0000	&	0.0000	\\
Self Energy	&	0.0000	&	0.0000	\\
Vac. Pol. (Uehling)	&	1.1678	&	1.8769	\\
Vac. Pol. Wichman-Kroll	&	-0.0007	&	-0.0008	\\
Vac. Pol. Loop after Loop	&	0.0007	&	0.0016	\\
Vac. Pol. Källén-Sabry	&	0.0111	&	0.0152	\\
Relativistic Recoil 	&	0.0025	&	0.0025	\\
HFS Shift	&	-0.0006	&	-0.0008	\\
Total	&	2969.6374	&	2970.4182	\\
\hline
Error & 0.0005 & 0.0005\\
Error due to the kaon mass & 0.096 & 0.096\\
\hline
\hline
\end{tabular}
\end{table} 
\begin{table} 
\caption{Hyperfine transition energies and transition rate in kaonic nitrogen.
\label{tab:kN_energies}
} 
\begin{tabular}{c c r r r r } 
\hline 
\hline
Transition	&	F-F'	&	Trans. rate (s$^{-1}$)	&	Trans. E (eV)	&	Shift (eV)	&		\\	
\hline
$8i \to 7h$	&	7-6	& $	1.19	\times 10^{13}$	&	2970.4169	&	-0.00216			\\
	&	6-5	& $	1.00	\times 10^{13}$	&	2970.4196	&	0.00050			\\
	&	6-6	& $	0.98	\times 10^{13}$	&	2970.4145	&	-0.00453			\\
	&	5-4	& $	0.84	\times 10^{13}$	&	2970.4217	&	0.00265			\\
	&	5-5	& $	0.03	\times 10^{13}$	&	2970.4175	&	-0.00154			\\
	&	5-6	& $	0.00	\times 10^{13}$	&	2970.4125	&	-0.00656			\\
$8k \to 7i$ 	&	8-7	& $	1.54	\times 10^{13}$	&	2969.6365	&	-0.00149			\\
	&	7-6	& $	1.33	\times 10^{13}$	&	2969.6383	&	0.00029			\\
	&	7-7	& $	1.31	\times 10^{13}$	&	2969.6347	&	-0.00326			\\
	&	6-5	& $	1.15	\times 10^{13}$	&	2969.6398	&	0.00178			\\
	&	6-6	& $	0.03	\times 10^{13}$	&	2969.6367	&	-0.00126			\\
	&	6-7	& $	0.00	\times 10^{13}$	&	2969.6332	&	-0.00480			\\
\hline 
\hline
\end{tabular} 
\end{table} 

The transition energies for the $8 \to 7$ transitions in kaonic nitrogen are presented in Tables~\ref{tab:kN} and \ref{tab:kN_energies}. As for the pionic nitrogen, the error contribution due to the QED correction not considered is dominated by the unevaluated diagram  
with the vacuum polarization loop in the nuclear photon line \cite{Pachucki1996},
the associated correction is estimated in the order of 0.5~meV.
For $8k \to 7i$ and $8i \to 7h$ transitions we have a HFS shift of 0.6 and 0.8~meV, respectively, which correspond to a correction of the kaon mass between 0.2 and 0.3~ppm.

As a general note, we remark that if we assume a statistical distribution of the initial state sublevels populations, transitions $n l \to n' s$ with a $s$ orbital as
 final state  have an average HFS  shift equal to zero due to an exact cancellation between the weighted excited sublevels energy shifts as seen
from  Eq.~\eqref{eq:trans-prob}.

\subsection{General behavior of the hyperfine structure correction over Z} 
For the  non-relativistic case, 
 the HFS splitting normalized to the 
binding energy and to the nuclear magnetic moment, depends linearly on $Z \alpha$.
Any deviation from this linear dependence in the Klein-Gordon HFS can be attributed only to
relativistic effects.

To study the behavior of the normalized HFS splitting  $(E^{9p}_{F=3/2}-E^{9p}_{F=3/2})/(\E_0 \mu_I)$ for the relativistic case,  
we calculated the HFS for a selected choice of 
pionic atoms with a stable nucleus of spin 1/2.  
The orbital $9p$ has been chosen to minimize the effect of the finite nuclear size and strong interaction shifts,
 particularly for high values of $Z$.
The results are
summarized in Table~\ref{tab:pions-scan12}.
For these calculations we used the nuclear mass values from Ref.~\cite{Audi2003},
the nuclear radii from Refs.~\cite{CODATA2002,Angeli2004} and the nuclear magnetic moments 
from Ref.~\cite{Raghavan1989}.
  
\begin{figure} 

 \includegraphics[width=0.45\textwidth]{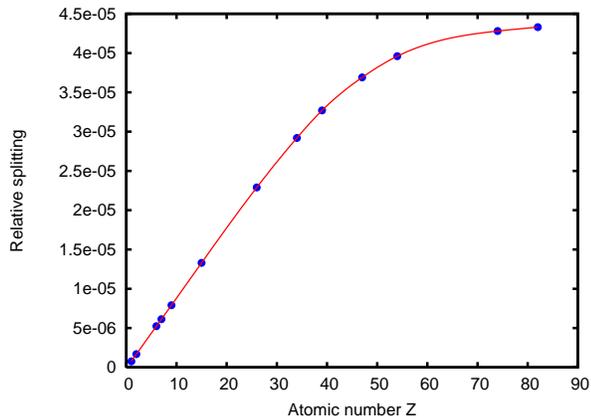}   
\caption{Value of relative splitting $(E^{9p}_{F=3/2}-E^{9p}_{F=3/2})/(\E_0 \mu_I)$ for pionic atoms with different values of $Z$. Nuclear masses, radii and magnetic moments values have been obtained from Refs.~\cite{Audi2003,Angeli2004,CODATA2002,Raghavan1989}.}
\label{pions-scan12}  
\end{figure}

For higher $Z$ values a non-linear dependence on $Z \alpha$ appears as we can see in  
Fig.~\ref{pions-scan12}. This non-linearity originates  in the two different parts of  
 Eq.~\eqref{eq:HFS_final_SI}: the non-trivial dependency on $\E_0$ in the denominator
and the expectation  value $\langle nl|  r^{-3} | n l\rangle$.

\begin{table} 
\caption{HFS separation of the F=1/2 and F=3/2  levels for the $9p$ orbital for pionic atoms with spin $\frac{1}{ 2}$ nucleus.
\label{tab:pions-scan12} 
}
\begin{tabular}{ l  r  r  r }
\hline
\hline
Element   &       Z &       $9p$ energy (eV)   &       HFS splitting (eV)      \\
\hline                                                  
$\mathrm{H}$	&	1	&	-39.93816	&	0.0001	\\
$\mathrm{^3He}$	&	2	&	-174.8370	&	-0.0009	\\
$\mathrm{^{13}C}$	&	6	&	-1633.402	&	0.0060	\\
$\mathrm{^{15}N}$	&	7	&	-2226.813	&	-0.0039	\\
$\mathrm{^{19}F}$	&	9	&	-3689.435	&	0.0767	\\
$\mathrm{^{31}P}$	&	15	&	-10286.63	&	0.1544	\\
$\mathrm{^{57}Fe}$	&	26	&	-31023.63	&	0.0643	\\
$\mathrm{^{77}Se}$	&	34	&	-53146.21	&	0.8293	\\
$\mathrm{^{89}Y}$	&	39	&	-69987.00	&	-0.3143	\\
$\mathrm{^{107}Ag}$	&	47	&	-101681.2	&	-0.4266	\\
$\mathrm{^{129}Xe}$	&	54	&	-134178.0	&	-4.1295	\\
$\mathrm{^{183}W}$	&	74	&	-250634.4	&	1.2629	\\
$\mathrm{^{202}Pb}$	&	82	&	-306731.6	&	7.8662	\\
\hline
\hline
\end{tabular}
\end{table} 

\section{Conclusions}
We presented a relativistic calculation of the hyperfine structure in pionic and kaonic atoms. 
The precise evaluation of the specific case of  pionic and kaonic nitrogen is
 particularly important for the new measurement of the pion and kaon mass. 
The small error on the theoretical predictions, of the order of 1~meV
 for the $5 \to 4$ transition, corresponds to a systematic error of $\gtrsim 0.2$~ppm for the pion
 mass evaluation, considerably smaller than  the error of
 previous theoretical predictions \cite{Schroder2001}.

The formalism presented in this article can be applied for other effect as the quadrupole nuclear moment which can not be negligible for mesonic atoms with high $Z$.
In this case, HFS due to the quadrupole moment can be predicted using the next  multipole in the 
development of the electric potential of the nucleus to evaluate the correspondent perturbation operator.  
This application is particularly important for the calculation of the atomic levels in heavy pionic ions, where relativistic and nucleus deformation effects can be taken into account at the same time.

\section*{Acknowledgement}
We thank B. Loiseau, T. Ericson, D. Gotta and L. Simons for interesting discussion about pionic atoms.
One of the author (M.T.) was partially sponsored by the Alexander von Humbouldt Foundation.
Laboratoire Kastler Brossel is Unité Mixte de Recherche du CNRS 
n$^{\circ}$ 8552.

\bibliography{martino}

\begin{thebibliography}{52}
\expandafter\ifx\csname natexlab\endcsname\relax\def\natexlab#1{#1}\fi
\expandafter\ifx\csname bibnamefont\endcsname\relax
  \def\bibnamefont#1{#1}\fi
\expandafter\ifx\csname bibfnamefont\endcsname\relax
  \def\bibfnamefont#1{#1}\fi
\expandafter\ifx\csname citenamefont\endcsname\relax
  \def\citenamefont#1{#1}\fi
\expandafter\ifx\csname url\endcsname\relax
  \def\url#1{\texttt{#1}}\fi
\expandafter\ifx\csname urlprefix\endcsname\relax\def\urlprefix{URL }\fi
\providecommand{\bibinfo}[2]{#2}
\providecommand{\eprint}[2][]{\url{#2}}

\bibitem[{\citenamefont{Gotta}(2004)}]{Gotta2004}
\bibinfo{author}{\bibfnamefont{D.}~\bibnamefont{Gotta}},
  \bibinfo{journal}{Progress in Particle and Nuclear Physics}
  \textbf{\bibinfo{volume}{52}}, \bibinfo{pages}{133} (\bibinfo{year}{2004}).

\bibitem[{\citenamefont{{Beer} et~al.}(2002)\citenamefont{{Beer},
  {Bragadireanu}, {Breunlich}, {Cargnelli}, {Curceanu (Petrascu)}, {Egger},
  {Fuhrmann}, {Guaraldo}, {Giersch}, {Iliescu} et~al.}}]{Beer2002}
\bibinfo{author}{\bibfnamefont{G.}~\bibnamefont{{Beer}}},
  \bibinfo{author}{\bibfnamefont{A.~M.} \bibnamefont{{Bragadireanu}}},
  \bibinfo{author}{\bibfnamefont{W.}~\bibnamefont{{Breunlich}}},
  \bibinfo{author}{\bibfnamefont{M.}~\bibnamefont{{Cargnelli}}},
  \bibinfo{author}{\bibfnamefont{C.}~\bibnamefont{{Curceanu (Petrascu)}}},
  \bibinfo{author}{\bibfnamefont{J.-P.} \bibnamefont{{Egger}}},
  \bibinfo{author}{\bibfnamefont{H.}~\bibnamefont{{Fuhrmann}}},
  \bibinfo{author}{\bibfnamefont{C.}~\bibnamefont{{Guaraldo}}},
  \bibinfo{author}{\bibfnamefont{M.}~\bibnamefont{{Giersch}}},
  \bibinfo{author}{\bibfnamefont{M.}~\bibnamefont{{Iliescu}}},
  \bibnamefont{et~al.}, \bibinfo{journal}{\plb} \textbf{\bibinfo{volume}{535}},
  \bibinfo{pages}{52} (\bibinfo{year}{2002}).

\bibitem[{\citenamefont{Gasser and Leutwyler}(1984)}]{Gasser1983}
\bibinfo{author}{\bibfnamefont{J.}~\bibnamefont{Gasser}} \bibnamefont{and}
  \bibinfo{author}{\bibfnamefont{H.}~\bibnamefont{Leutwyler}},
  \bibinfo{journal}{\ap} \textbf{\bibinfo{volume}{158}}, \bibinfo{pages}{142}
  (\bibinfo{year}{1984}).

\bibitem[{\citenamefont{Lyubovitskij and Rusetsky}(2000)}]{Lyubovitskij2000}
\bibinfo{author}{\bibfnamefont{V.~E.} \bibnamefont{Lyubovitskij}}
  \bibnamefont{and} \bibinfo{author}{\bibfnamefont{A.}~\bibnamefont{Rusetsky}},
  \bibinfo{journal}{\plb} \textbf{\bibinfo{volume}{494}}, \bibinfo{pages}{9}
  (\bibinfo{year}{2000}).

\bibitem[{\citenamefont{Gasser et~al.}(2002)\citenamefont{Gasser, Ivanov,
  Lipartia, Mojzis, and Rusetsky}}]{Gasser2002}
\bibinfo{author}{\bibfnamefont{J.}~\bibnamefont{Gasser}},
  \bibinfo{author}{\bibfnamefont{M.~A.} \bibnamefont{Ivanov}},
  \bibinfo{author}{\bibfnamefont{E.}~\bibnamefont{Lipartia}},
  \bibinfo{author}{\bibfnamefont{M.}~\bibnamefont{Mojzis}}, \bibnamefont{and}
  \bibinfo{author}{\bibfnamefont{A.}~\bibnamefont{Rusetsky}},
  \bibinfo{journal}{\epjc} \textbf{\bibinfo{volume}{26}}, \bibinfo{pages}{13}
  (\bibinfo{year}{2002}), \eprint{hep-ph/0206068}.

\bibitem[{\citenamefont{{Pionic Hydrogen Collaboration}}(1998)}]{ProposalPH}
\bibinfo{author}{\bibnamefont{{Pionic Hydrogen Collaboration}}},
  \bibinfo{journal}{PSI experiment proposal R-98.01}  (\bibinfo{year}{1998}),
  \urlprefix\url{http://pihydrogen.web.psi.ch}.

\bibitem[{\citenamefont{Anagnostopoulos
  et~al.}(2003{\natexlab{a}})\citenamefont{Anagnostopoulos, Cargnelli,
  Fuhrmann, Giersch, Gotta, Gruber, Hennebach, Hirtl, Indelicato, Liu
  et~al.}}]{Anagnostopoulos2003a}
\bibinfo{author}{\bibfnamefont{D.~F.} \bibnamefont{Anagnostopoulos}},
  \bibinfo{author}{\bibfnamefont{M.}~\bibnamefont{Cargnelli}},
  \bibinfo{author}{\bibfnamefont{H.}~\bibnamefont{Fuhrmann}},
  \bibinfo{author}{\bibfnamefont{M.}~\bibnamefont{Giersch}},
  \bibinfo{author}{\bibfnamefont{D.}~\bibnamefont{Gotta}},
  \bibinfo{author}{\bibfnamefont{A.}~\bibnamefont{Gruber}},
  \bibinfo{author}{\bibfnamefont{M.}~\bibnamefont{Hennebach}},
  \bibinfo{author}{\bibfnamefont{A.}~\bibnamefont{Hirtl}},
  \bibinfo{author}{\bibfnamefont{P.}~\bibnamefont{Indelicato}},
  \bibinfo{author}{\bibfnamefont{Y.~W.} \bibnamefont{Liu}},
  \bibnamefont{et~al.}, \bibinfo{journal}{\npa} \textbf{\bibinfo{volume}{721}},
  \bibinfo{pages}{849c} (\bibinfo{year}{2003}{\natexlab{a}}).

\bibitem[{\citenamefont{Beer et~al.}(2005)\citenamefont{Beer, Bragadireanu,
  Cargnelli, Curceanu-Petrascu, Egger, Fuhrmann, Guaraldo, Iliescu, Ishiwatari,
  Itahashi et~al.}}]{Beer2005}
\bibinfo{author}{\bibfnamefont{G.}~\bibnamefont{Beer}},
  \bibinfo{author}{\bibfnamefont{A.~M.} \bibnamefont{Bragadireanu}},
  \bibinfo{author}{\bibfnamefont{M.}~\bibnamefont{Cargnelli}},
  \bibinfo{author}{\bibfnamefont{C.}~\bibnamefont{Curceanu-Petrascu}},
  \bibinfo{author}{\bibfnamefont{J.-P.} \bibnamefont{Egger}},
  \bibinfo{author}{\bibfnamefont{H.}~\bibnamefont{Fuhrmann}},
  \bibinfo{author}{\bibfnamefont{C.}~\bibnamefont{Guaraldo}},
  \bibinfo{author}{\bibfnamefont{M.}~\bibnamefont{Iliescu}},
  \bibinfo{author}{\bibfnamefont{T.}~\bibnamefont{Ishiwatari}},
  \bibinfo{author}{\bibfnamefont{K.}~\bibnamefont{Itahashi}},
  \bibnamefont{et~al.} (\bibinfo{collaboration}{DEAR Collaboration}),
  \bibinfo{journal}{\prl} \textbf{\bibinfo{volume}{94}}, \bibinfo{eid}{212302}
  (pages~\bibinfo{numpages}{4}) (\bibinfo{year}{2005}).

\bibitem[{\citenamefont{Anagnostopoulos
  et~al.}(2003{\natexlab{b}})\citenamefont{Anagnostopoulos, Gotta, Indelicato,
  and Simons}}]{Anagnostopoulos2003b}
\bibinfo{author}{\bibfnamefont{D.~F.} \bibnamefont{Anagnostopoulos}},
  \bibinfo{author}{\bibfnamefont{D.}~\bibnamefont{Gotta}},
  \bibinfo{author}{\bibfnamefont{P.}~\bibnamefont{Indelicato}},
  \bibnamefont{and} \bibinfo{author}{\bibfnamefont{L.~M.}
  \bibnamefont{Simons}}, \bibinfo{journal}{\prl} \textbf{\bibinfo{volume}{91}},
  \bibinfo{pages}{240801} (\bibinfo{year}{2003}{\natexlab{b}}).

\bibitem[{\citenamefont{Lenz et~al.}(1998)\citenamefont{Lenz, Borchert, Gorke,
  Gotta, Siems, Anagnostopoulos, Augsburger, Chatellard, Egger, Belmiloud
  et~al.}}]{Lenz1998}
\bibinfo{author}{\bibfnamefont{S.}~\bibnamefont{Lenz}},
  \bibinfo{author}{\bibfnamefont{G.}~\bibnamefont{Borchert}},
  \bibinfo{author}{\bibfnamefont{H.}~\bibnamefont{Gorke}},
  \bibinfo{author}{\bibfnamefont{D.}~\bibnamefont{Gotta}},
  \bibinfo{author}{\bibfnamefont{T.}~\bibnamefont{Siems}},
  \bibinfo{author}{\bibfnamefont{D.~F.} \bibnamefont{Anagnostopoulos}},
  \bibinfo{author}{\bibfnamefont{M.}~\bibnamefont{Augsburger}},
  \bibinfo{author}{\bibfnamefont{D.}~\bibnamefont{Chatellard}},
  \bibinfo{author}{\bibfnamefont{J.~P.} \bibnamefont{Egger}},
  \bibinfo{author}{\bibfnamefont{D.}~\bibnamefont{Belmiloud}},
  \bibnamefont{et~al.}, \bibinfo{journal}{\plb} \textbf{\bibinfo{volume}{416}},
  \bibinfo{pages}{50} (\bibinfo{year}{1998}).

\bibitem[{\citenamefont{{Pion Mass Collaboration}}(1997)}]{ProposalPM}
\bibinfo{author}{\bibnamefont{{Pion Mass Collaboration}}},
  \bibinfo{journal}{PSI experiment proposal R-97.02}  (\bibinfo{year}{1997}).

\bibitem[{\citenamefont{Nelms et~al.}(2002)\citenamefont{Nelms,
  Anagnostopoulos, Augsburger, Borchert, Chatellard, Daum, Egger, Gotta,
  Hauser, Indelicato et~al.}}]{Nelms2002a}
\bibinfo{author}{\bibfnamefont{N.}~\bibnamefont{Nelms}},
  \bibinfo{author}{\bibfnamefont{D.~F.} \bibnamefont{Anagnostopoulos}},
  \bibinfo{author}{\bibfnamefont{M.}~\bibnamefont{Augsburger}},
  \bibinfo{author}{\bibfnamefont{G.}~\bibnamefont{Borchert}},
  \bibinfo{author}{\bibfnamefont{D.}~\bibnamefont{Chatellard}},
  \bibinfo{author}{\bibfnamefont{M.}~\bibnamefont{Daum}},
  \bibinfo{author}{\bibfnamefont{J.~P.} \bibnamefont{Egger}},
  \bibinfo{author}{\bibfnamefont{D.}~\bibnamefont{Gotta}},
  \bibinfo{author}{\bibfnamefont{P.}~\bibnamefont{Hauser}},
  \bibinfo{author}{\bibfnamefont{P.}~\bibnamefont{Indelicato}},
  \bibnamefont{et~al.}, \bibinfo{journal}{\nima}
  \textbf{\bibinfo{volume}{477}}, \bibinfo{pages}{461} (\bibinfo{year}{2002}).

\bibitem[{\citenamefont{Trassinelli}(2005)}]{MyThesis}
\bibinfo{author}{\bibfnamefont{M.}~\bibnamefont{Trassinelli}}, Ph.D. thesis,
  \bibinfo{school}{Université Pierre et Marie Curie}, \bibinfo{address}{Paris,
  France} (\bibinfo{year}{2005}),
  \urlprefix\url{http://tel.ccsd.cnrs.fr/tel-00067768}.

\bibitem[{\citenamefont{Ebersold et~al.}(1978)\citenamefont{Ebersold, Aas, Dey,
  Eichler, Leisi, Sapp, and Scheck}}]{Ebersold1978}
\bibinfo{author}{\bibfnamefont{P.}~\bibnamefont{Ebersold}},
  \bibinfo{author}{\bibfnamefont{B.}~\bibnamefont{Aas}},
  \bibinfo{author}{\bibfnamefont{W.}~\bibnamefont{Dey}},
  \bibinfo{author}{\bibfnamefont{R.}~\bibnamefont{Eichler}},
  \bibinfo{author}{\bibfnamefont{H.~J.} \bibnamefont{Leisi}},
  \bibinfo{author}{\bibfnamefont{W.~W.} \bibnamefont{Sapp}}, \bibnamefont{and}
  \bibinfo{author}{\bibfnamefont{F.}~\bibnamefont{Scheck}},
  \bibinfo{journal}{\npa} \textbf{\bibinfo{volume}{296}}, \bibinfo{pages}{493}
  (\bibinfo{year}{1978}).

\bibitem[{\citenamefont{Koch and Scheck}(1980)}]{Koch1980}
\bibinfo{author}{\bibfnamefont{J.}~\bibnamefont{Koch}} \bibnamefont{and}
  \bibinfo{author}{\bibfnamefont{F.}~\bibnamefont{Scheck}},
  \bibinfo{journal}{\npa} \textbf{\bibinfo{volume}{340}}, \bibinfo{pages}{221}
  (\bibinfo{year}{1980}).

\bibitem[{\citenamefont{Scheck}(1983)}]{Scheck}
\bibinfo{author}{\bibfnamefont{F.}~\bibnamefont{Scheck}},
  \emph{\bibinfo{title}{Leptons, Hadrons, and Nuclei}}
  (\bibinfo{publisher}{Elsevier}, \bibinfo{address}{North-Holland},
  \bibinfo{year}{1983}), \bibinfo{edition}{1st} ed.

\bibitem[{\citenamefont{Austen and de~Swart}(1983)}]{Austen1983}
\bibinfo{author}{\bibfnamefont{G.}~\bibnamefont{Austen}} \bibnamefont{and}
  \bibinfo{author}{\bibfnamefont{J.}~\bibnamefont{de~Swart}},
  \bibinfo{journal}{\prl} \textbf{\bibinfo{volume}{50}}, \bibinfo{pages}{2039}
  (\bibinfo{year}{1983}).

\bibitem[{\citenamefont{Owen}(1994)}]{Owen1994}
\bibinfo{author}{\bibfnamefont{D.~A.} \bibnamefont{Owen}},
  \bibinfo{journal}{\fp} \textbf{\bibinfo{volume}{24}}, \bibinfo{pages}{273}
  (\bibinfo{year}{1994}).

\bibitem[{\citenamefont{Trassinelli et~al.}(2006)\citenamefont{Trassinelli,
  Anagnostopoulos, Borchert, A., Dax, Egger, Gotta, Hennebach, Indelicato, Liu
  et~al.}}]{Trassinelli2007}
\bibinfo{author}{\bibfnamefont{M.}~\bibnamefont{Trassinelli}},
  \bibinfo{author}{\bibfnamefont{D.~F.} \bibnamefont{Anagnostopoulos}},
  \bibinfo{author}{\bibfnamefont{G.}~\bibnamefont{Borchert}},
  \bibinfo{author}{\bibnamefont{A.}}, \bibinfo{author}{\bibnamefont{Dax}},
  \bibinfo{author}{\bibfnamefont{J.~P.} \bibnamefont{Egger}},
  \bibinfo{author}{\bibfnamefont{D.}~\bibnamefont{Gotta}},
  \bibinfo{author}{\bibfnamefont{M.}~\bibnamefont{Hennebach}},
  \bibinfo{author}{\bibfnamefont{P.}~\bibnamefont{Indelicato}},
  \bibinfo{author}{\bibfnamefont{Y.-W.} \bibnamefont{Liu}},
  \bibnamefont{et~al.} (\bibinfo{year}{2006}), \bibinfo{note}{in preparation}.

\bibitem[{\citenamefont{Bjorken and Drell}(1964)}]{Bjorken-Drell-RQM}
\bibinfo{author}{\bibfnamefont{J.}~\bibnamefont{Bjorken}} \bibnamefont{and}
  \bibinfo{author}{\bibfnamefont{S.}~\bibnamefont{Drell}},
  \emph{\bibinfo{title}{Relativistic Quantum Mechanics}}
  (\bibinfo{publisher}{McGraw-Hill Book Company}, \bibinfo{address}{San
  Francisco}, \bibinfo{year}{1964}), \bibinfo{edition}{1st} ed.

\bibitem[{\citenamefont{Decker et~al.}(1987)\citenamefont{Decker, Pilkuhn, and
  Schlageter}}]{Decker1987}
\bibinfo{author}{\bibfnamefont{R.}~\bibnamefont{Decker}},
  \bibinfo{author}{\bibfnamefont{H.}~\bibnamefont{Pilkuhn}}, \bibnamefont{and}
  \bibinfo{author}{\bibfnamefont{A.}~\bibnamefont{Schlageter}},
  \bibinfo{journal}{\zphd} \textbf{\bibinfo{volume}{6}}, \bibinfo{pages}{1}
  (\bibinfo{year}{1987}).

\bibitem[{\citenamefont{Lee et~al.}(2006)\citenamefont{Lee, Milstein, and
  Karshenboim}}]{Lee2006}
\bibinfo{author}{\bibfnamefont{R.~N.} \bibnamefont{Lee}},
  \bibinfo{author}{\bibfnamefont{A.~I.} \bibnamefont{Milstein}},
  \bibnamefont{and} \bibinfo{author}{\bibfnamefont{S.~G.}
  \bibnamefont{Karshenboim}}, \bibinfo{journal}{\pra}
  \textbf{\bibinfo{volume}{73}}, \bibinfo{eid}{012505}
  (pages~\bibinfo{numpages}{4}) (\bibinfo{year}{2006}).

\bibitem[{\citenamefont{Friar}(1980)}]{Friar1980}
\bibinfo{author}{\bibfnamefont{J.~L.} \bibnamefont{Friar}},
  \bibinfo{journal}{\zph} \textbf{\bibinfo{volume}{297}}, \bibinfo{pages}{147}
  (\bibinfo{year}{1980}).

\bibitem[{\citenamefont{{Leon} and {Seki}}(1981)}]{Leon1981}
\bibinfo{author}{\bibfnamefont{M.}~\bibnamefont{{Leon}}} \bibnamefont{and}
  \bibinfo{author}{\bibfnamefont{R.}~\bibnamefont{{Seki}}},
  \bibinfo{journal}{\npa} \textbf{\bibinfo{volume}{352}}, \bibinfo{pages}{477}
  (\bibinfo{year}{1981}).

\bibitem[{\citenamefont{Schwartz}(1955)}]{Schwartz1955}
\bibinfo{author}{\bibfnamefont{C.}~\bibnamefont{Schwartz}},
  \bibinfo{journal}{\pr} \textbf{\bibinfo{volume}{97}}, \bibinfo{pages}{380}
  (\bibinfo{year}{1955}).

\bibitem[{\citenamefont{Lindgren and Rosén}(1974)}]{Lindgren1974}
\bibinfo{author}{\bibfnamefont{I.}~\bibnamefont{Lindgren}} \bibnamefont{and}
  \bibinfo{author}{\bibfnamefont{A.}~\bibnamefont{Rosén}},
  \bibinfo{journal}{Case Studies in Atomic Physics}
  \textbf{\bibinfo{volume}{4}}, \bibinfo{pages}{93} (\bibinfo{year}{1974}).

\bibitem[{\citenamefont{Cheng and Childs}(1985)}]{Cheng1985}
\bibinfo{author}{\bibfnamefont{K.}~\bibnamefont{Cheng}} \bibnamefont{and}
  \bibinfo{author}{\bibfnamefont{W.}~\bibnamefont{Childs}},
  \bibinfo{journal}{\pra} \textbf{\bibinfo{volume}{31}}, \bibinfo{pages}{2775}
  (\bibinfo{year}{1985}).

\bibitem[{\citenamefont{Bohr and Weisskopf}(1950)}]{Bohr1950}
\bibinfo{author}{\bibfnamefont{A.}~\bibnamefont{Bohr}} \bibnamefont{and}
  \bibinfo{author}{\bibfnamefont{V.~F.} \bibnamefont{Weisskopf}},
  \bibinfo{journal}{\pr} \textbf{\bibinfo{volume}{77}}, \bibinfo{pages}{94}
  (\bibinfo{year}{1950}).

\bibitem[{\citenamefont{Edmonds}(1974)}]{Edmonds}
\bibinfo{author}{\bibfnamefont{A.~R.} \bibnamefont{Edmonds}},
  \emph{\bibinfo{title}{Angular Momentum in {Q}uantum {M}echanics}}
  (\bibinfo{publisher}{Princeton University Press}, \bibinfo{year}{1974}),
  \bibinfo{edition}{3rd} ed.

\bibitem[{\citenamefont{Lindgren and Morrison}(1982)}]{Lindgren-Morrison}
\bibinfo{author}{\bibfnamefont{I.}~\bibnamefont{Lindgren}} \bibnamefont{and}
  \bibinfo{author}{\bibfnamefont{J.}~\bibnamefont{Morrison}},
  \emph{\bibinfo{title}{Atomic Many-Body Theory}}, Atoms and Plasmas
  (\bibinfo{publisher}{Springer}, \bibinfo{address}{Berlin},
  \bibinfo{year}{1982}), \bibinfo{edition}{2nd} ed.

\bibitem[{\citenamefont{Varshalovich et~al.}(1988)\citenamefont{Varshalovich,
  Moskalev, and Khersonskii}}]{Varshalovich}
\bibinfo{author}{\bibfnamefont{D.~A.} \bibnamefont{Varshalovich}},
  \bibinfo{author}{\bibfnamefont{A.~N.} \bibnamefont{Moskalev}},
  \bibnamefont{and} \bibinfo{author}{\bibfnamefont{V.~K.}
  \bibnamefont{Khersonskii}}, \emph{\bibinfo{title}{Quantum Theory of Angular
  Momentum}} (\bibinfo{publisher}{World Scientific},
  \bibinfo{address}{Singapore}, \bibinfo{year}{1988}), \bibinfo{edition}{1st}
  ed.

\bibitem[{\citenamefont{Bethe and Salpeter}(1957)}]{Bethe-Salpeter}
\bibinfo{author}{\bibfnamefont{H.~B.} \bibnamefont{Bethe}} \bibnamefont{and}
  \bibinfo{author}{\bibfnamefont{E.~E.} \bibnamefont{Salpeter}},
  \emph{\bibinfo{title}{Quantum Mechanics of One- and Two-Electron Atoms}}
  (\bibinfo{publisher}{Springer-Verlag}, \bibinfo{year}{1957}),
  \bibinfo{edition}{1st} ed.

\bibitem[{\citenamefont{Desclaux}(1975)}]{Desclaux1975}
\bibinfo{author}{\bibfnamefont{J.~P.} \bibnamefont{Desclaux}},
  \bibinfo{journal}{\cpc} \textbf{\bibinfo{volume}{9}}, \bibinfo{pages}{31}
  (\bibinfo{year}{1975}).

\bibitem[{\citenamefont{Desclaux}(1993)}]{Desclaux1993}
\bibinfo{author}{\bibfnamefont{J.~P.} \bibnamefont{Desclaux}}, in
  \emph{\bibinfo{booktitle}{Methods and Techniques in Computational
  Chemistry}}, edited by
  \bibinfo{editor}{\bibfnamefont{E.}~\bibnamefont{Clementi}}
  (\bibinfo{publisher}{STEF}, \bibinfo{address}{Cagliary},
  \bibinfo{year}{1993}), vol. \bibinfo{volume}{A: Small Systems} of
  \emph{\bibinfo{series}{METTEC}}, p. \bibinfo{pages}{253},
  \urlprefix\url{hhtp://dirac.spectro.jussieu.fr/mcdf}.

\bibitem[{\citenamefont{Boucard and Indelicato}(2000)}]{Boucard2000}
\bibinfo{author}{\bibfnamefont{S.}~\bibnamefont{Boucard}} \bibnamefont{and}
  \bibinfo{author}{\bibfnamefont{P.}~\bibnamefont{Indelicato}},
  \bibinfo{journal}{\epjd} \textbf{\bibinfo{volume}{8}}, \bibinfo{pages}{59}
  (\bibinfo{year}{2000}).

\bibitem[{\citenamefont{Desclaux et~al.}(2003)\citenamefont{Desclaux,
  Dolbeault, Esteban, Indelicato, and Séré}}]{Desclaux2003}
\bibinfo{author}{\bibfnamefont{J.}~\bibnamefont{Desclaux}},
  \bibinfo{author}{\bibfnamefont{J.}~\bibnamefont{Dolbeault}},
  \bibinfo{author}{\bibfnamefont{M.}~\bibnamefont{Esteban}},
  \bibinfo{author}{\bibfnamefont{P.}~\bibnamefont{Indelicato}},
  \bibnamefont{and} \bibinfo{author}{\bibfnamefont{E.}~\bibnamefont{Séré}}, in
  \emph{\bibinfo{booktitle}{Computational Chemistry}}, edited by
  \bibinfo{editor}{\bibfnamefont{P.}~\bibnamefont{Ciarlet}}
  (\bibinfo{publisher}{Elsevier}, \bibinfo{year}{2003}),
  vol.~\bibinfo{volume}{X}, p. \bibinfo{pages}{1032}.

\bibitem[{\citenamefont{Santos et~al.}(2005)\citenamefont{Santos, Parente,
  Boucard, Indelicato, and Desclaux}}]{Santos2005}
\bibinfo{author}{\bibfnamefont{J.~P.} \bibnamefont{Santos}},
  \bibinfo{author}{\bibfnamefont{F.}~\bibnamefont{Parente}},
  \bibinfo{author}{\bibfnamefont{S.}~\bibnamefont{Boucard}},
  \bibinfo{author}{\bibfnamefont{P.}~\bibnamefont{Indelicato}},
  \bibnamefont{and} \bibinfo{author}{\bibfnamefont{J.~P.}
  \bibnamefont{Desclaux}}, \bibinfo{journal}{\pra}
  \textbf{\bibinfo{volume}{71}}, \bibinfo{eid}{032501}
  (pages~\bibinfo{numpages}{8}) (\bibinfo{year}{2005}).

\bibitem[{\citenamefont{Béretetski et~al.}(1989)\citenamefont{Béretetski,
  Lifchitz, and Pitayevsky}}]{LandauEMQ}
\bibinfo{author}{\bibfnamefont{V.}~\bibnamefont{Béretetski}},
  \bibinfo{author}{\bibfnamefont{E.}~\bibnamefont{Lifchitz}}, \bibnamefont{and}
  \bibinfo{author}{\bibfnamefont{L.}~\bibnamefont{Pitayevsky}},
  \emph{\bibinfo{title}{Électrodynamique Quantique}}, Physique Theorique
  (\bibinfo{publisher}{Éditions MIR}, \bibinfo{address}{Moscou},
  \bibinfo{year}{1989}), \bibinfo{edition}{2nd} ed.

\bibitem[{\citenamefont{Audi et~al.}(2003)\citenamefont{Audi, Wapstra, and
  Thibault}}]{Audi2003}
\bibinfo{author}{\bibfnamefont{G.}~\bibnamefont{Audi}},
  \bibinfo{author}{\bibfnamefont{A.}~\bibnamefont{Wapstra}}, \bibnamefont{and}
  \bibinfo{author}{\bibfnamefont{C.}~\bibnamefont{Thibault}},
  \bibinfo{journal}{\npa} \textbf{\bibinfo{volume}{729}}, \bibinfo{pages}{337}
  (\bibinfo{year}{2003}).

\bibitem[{\citenamefont{Boucard}(2000)}]{BoucardThesis}
\bibinfo{author}{\bibfnamefont{S.}~\bibnamefont{Boucard}}, Ph.D. thesis,
  \bibinfo{school}{Université Pierre et Marie Curie}, \bibinfo{address}{Paris,
  France} (\bibinfo{year}{2000}),
  \urlprefix\url{http://tel.ccsd.cnrs.fr/tel-00067768}.

\bibitem[{\citenamefont{{Eidelman} et~al.}(2004)\citenamefont{{Eidelman},
  {Hayes}, {Olive}, {Aguilar-Benitez}, {Amsler}, {Asner}, {Babu}, {Barnett},
  {Beringer}, {Burchat} et~al.}}]{PDBook}
\bibinfo{author}{\bibfnamefont{S.}~\bibnamefont{{Eidelman}}},
  \bibinfo{author}{\bibfnamefont{K.}~\bibnamefont{{Hayes}}},
  \bibinfo{author}{\bibfnamefont{K.}~\bibnamefont{{Olive}}},
  \bibinfo{author}{\bibfnamefont{M.}~\bibnamefont{{Aguilar-Benitez}}},
  \bibinfo{author}{\bibfnamefont{C.}~\bibnamefont{{Amsler}}},
  \bibinfo{author}{\bibfnamefont{D.}~\bibnamefont{{Asner}}},
  \bibinfo{author}{\bibfnamefont{K.}~\bibnamefont{{Babu}}},
  \bibinfo{author}{\bibfnamefont{R.}~\bibnamefont{{Barnett}}},
  \bibinfo{author}{\bibfnamefont{J.}~\bibnamefont{{Beringer}}},
  \bibinfo{author}{\bibfnamefont{P.}~\bibnamefont{{Burchat}}},
  \bibnamefont{et~al.}, \bibinfo{journal}{{\plb}}
  \textbf{\bibinfo{volume}{592}}, \bibinfo{pages}{1+} (\bibinfo{year}{2004}),
  \urlprefix\url{http://pdg.lbl.gov}.

\bibitem[{\citenamefont{Angeli}(2004)}]{Angeli2004}
\bibinfo{author}{\bibfnamefont{I.}~\bibnamefont{Angeli}},
  \bibinfo{journal}{\adndt} \textbf{\bibinfo{volume}{87}}, \bibinfo{pages}{185}
  (\bibinfo{year}{2004}).

\bibitem[{\citenamefont{Fullerton and Rinker}(1976)}]{Fullerton1976}
\bibinfo{author}{\bibfnamefont{L.~W.} \bibnamefont{Fullerton}}
  \bibnamefont{and} \bibinfo{author}{\bibfnamefont{G.~A.}
  \bibnamefont{Rinker}}, \bibinfo{journal}{\pra} \textbf{\bibinfo{volume}{13}},
  \bibinfo{pages}{1283} (\bibinfo{year}{1976}).

\bibitem[{\citenamefont{Huang}(1976)}]{Huang1976}
\bibinfo{author}{\bibfnamefont{K.~N.} \bibnamefont{Huang}},
  \bibinfo{journal}{\pra} \textbf{\bibinfo{volume}{14}}, \bibinfo{pages}{1311}
  (\bibinfo{year}{1976}).

\bibitem[{\citenamefont{Jeckelmann}(1985)}]{Jeckelmann1985}
\bibinfo{author}{\bibfnamefont{B.}~\bibnamefont{Jeckelmann}},
  \bibinfo{type}{Tech. Rep.}, \bibinfo{institution}{ETHZ-IMP}
  (\bibinfo{year}{1985}), \bibinfo{note}{lB-85-03}.

\bibitem[{\citenamefont{Barker and Glover}(1955)}]{Barker1955}
\bibinfo{author}{\bibfnamefont{W.~A.} \bibnamefont{Barker}} \bibnamefont{and}
  \bibinfo{author}{\bibfnamefont{F.~N.} \bibnamefont{Glover}},
  \bibinfo{journal}{\pr} \textbf{\bibinfo{volume}{99}}, \bibinfo{pages}{317}
  (\bibinfo{year}{1955}).

\bibitem[{\citenamefont{Pachucki}(1996)}]{Pachucki1996}
\bibinfo{author}{\bibfnamefont{K.}~\bibnamefont{Pachucki}},
  \bibinfo{journal}{\pra} \textbf{\bibinfo{volume}{53}}, \bibinfo{pages}{2092}
  (\bibinfo{year}{1996}).

\bibitem[{\citenamefont{Eides et~al.}(2001)\citenamefont{Eides, Grotch, and
  Shelyuto}}]{Eides2001}
\bibinfo{author}{\bibfnamefont{M.~I.} \bibnamefont{Eides}},
  \bibinfo{author}{\bibfnamefont{H.}~\bibnamefont{Grotch}}, \bibnamefont{and}
  \bibinfo{author}{\bibfnamefont{V.~A.} \bibnamefont{Shelyuto}},
  \bibinfo{journal}{\prep} \textbf{\bibinfo{volume}{342}}, \bibinfo{pages}{63}
  (\bibinfo{year}{2001}).

\bibitem[{\citenamefont{Sapirstein and Yennie}(1990)}]{Sapirstein1990}
\bibinfo{author}{\bibfnamefont{J.~R.} \bibnamefont{Sapirstein}}
  \bibnamefont{and} \bibinfo{author}{\bibfnamefont{D.~R.}
  \bibnamefont{Yennie}}, in \emph{\bibinfo{booktitle}{{Q}uantum
  {E}lectrodynamis}}, edited by
  \bibinfo{editor}{\bibfnamefont{T.}~\bibnamefont{Kinoshita}}
  (\bibinfo{publisher}{World Scientific}, \bibinfo{year}{1990}),
  vol.~\bibinfo{volume}{7} of \emph{\bibinfo{series}{Directions in High Energy
  Physics}}, p. \bibinfo{pages}{560}.

\bibitem[{\citenamefont{Mohr and Taylor}(2005)}]{CODATA2002}
\bibinfo{author}{\bibfnamefont{P.~J.} \bibnamefont{Mohr}} \bibnamefont{and}
  \bibinfo{author}{\bibfnamefont{B.~N.} \bibnamefont{Taylor}},
  \bibinfo{journal}{\rmp} \textbf{\bibinfo{volume}{77}}, \bibinfo{pages}{1}
  (\bibinfo{year}{2005}).

\bibitem[{\citenamefont{Raghavan}(1989)}]{Raghavan1989}
\bibinfo{author}{\bibfnamefont{P.}~\bibnamefont{Raghavan}},
  \bibinfo{journal}{\adndt} \textbf{\bibinfo{volume}{42}}, \bibinfo{pages}{189}
  (\bibinfo{year}{1989}).

\bibitem[{\citenamefont{Schröder et~al.}(2001)\citenamefont{Schröder,
  Badertscher, Goudsmit, Janousch, Leisi, Matsinos, Sigg, Zhao, Chatellard,
  Egger et~al.}}]{Schroder2001}
\bibinfo{author}{\bibfnamefont{H.~C.} \bibnamefont{Schröder}},
  \bibinfo{author}{\bibfnamefont{A.}~\bibnamefont{Badertscher}},
  \bibinfo{author}{\bibfnamefont{P.~F.~A.} \bibnamefont{Goudsmit}},
  \bibinfo{author}{\bibfnamefont{M.}~\bibnamefont{Janousch}},
  \bibinfo{author}{\bibfnamefont{H.}~\bibnamefont{Leisi}},
  \bibinfo{author}{\bibfnamefont{E.}~\bibnamefont{Matsinos}},
  \bibinfo{author}{\bibfnamefont{D.}~\bibnamefont{Sigg}},
  \bibinfo{author}{\bibfnamefont{Z.~G.} \bibnamefont{Zhao}},
  \bibinfo{author}{\bibfnamefont{D.}~\bibnamefont{Chatellard}},
  \bibinfo{author}{\bibfnamefont{J.-P.} \bibnamefont{Egger}},
  \bibnamefont{et~al.}, \bibinfo{journal}{\epjc} \textbf{\bibinfo{volume}{21}},
  \bibinfo{pages}{473} (\bibinfo{year}{2001}).

\end{thebibliography}

\end{document}